\documentclass[12pt,A4]{article}
\usepackage{graphics}
\begin{document}

\begin{center}
{\large \bf INVESTIGATION OF SOME OF THE THERMODYNAMICAL QUANTITIES OF PROTEINS BY STATISTICAL MECHANICAL METHODS}\\

\vspace{0.5 cm}
{\large G. Oylumluoglu$^{*1,2}$, Fevzi B\"{u}y\"{u}kk\i l\i \c{c}$^{**1}$,  Dogan Demirhan}$^{***1}$\\

$^{1}$Ege University, Faculty of Science, Department of
Physics, Izmir-TURKEY.\\

$^{2}$Mugla University, Faculty of Arts and
Sciences, Department of Physics, Mugla-TURKEY.\\

\end{center}

%--------------------------

\begin{abstract}
In this study, variations with respect to temperature of the
increments of enthalpy $\Delta H$ and Gibbs energy $\Delta G$,
arising in the dissolution of proteins in water, have been
investigated by the methods of statistical thermodynamics. In this
formalism, effective electric field $E$ and total dipole moment
$M$ are taken as thermodynamical variable. In obtaining the free
energy the partition function given by A. Bakk, J.S. Hoye and A.
Hansen; Physica A, 304, (2002), 355-361 has been used in a
modified form. In the constructed semi-phenomenological theory,
the experimental data are taken from the study of Privalov
\cite{{Privalov},{Privalov2}} and the relevant parameters have
been found by fitting to the experimental curves. The variations
of the increments of enthalpy  $\Delta H$ and Gibbs energy $\Delta
G$ have been investigated in the temperature range $265-350K$.

\vspace{0.5 cm} {\small Keywords: Thermodynamics, Classical
statistical mechanics}


\vspace{0.5 cm} {\small $^*$ Corresponding
Authors:e-mail:oylum@sci.ege.edu.tr, Phone:+90 232-3881892
(ext.2363)
Fax:+90 232-3881892\\
\\
} \vspace{0.5 cm} {\small $^{**}$ e-mail:fevzi@sci.ege.edu.tr,
Phone:+90 232-3881892 (ext.2846)
\\
} \vspace{0.5 cm} {\small $^{***}$ e-mail:dogan@sci.ege.edu.tr,
Phone:+90 232-3881892 (ext.2381)
\\
}
\end{abstract}

\section{Introduction}
\label{intro} Proteins is the common name of the complex macro
molecules which are formed by gathering of a great number
amino-acids and which play unremitting roles in the life times of
all of the living creatures. There are approximately twenty
different amino-acids which take place in the structure of the
proteins and which are chain molecules. The order or arrangement
of this twenty units is sensitive and definite. This order,
determines the nature of the protein and as a result identifies
its function \cite{Creighton}. Each protein molecule is formed by
on original combination and consecutive placing of different
numbers of amino-acids.

Proteins under their own natural physiological conditions,
spontaneously open out from one dimensional structures to three
dimensional structures in the time intervals starting from
milliseconds and getting to minutes \cite{{Hensen},{Hensen2}}. The
three dimensional structures of the proteins depend on various
environmental factors in the cell. Unfolded proteins return to
their natural states when the environmental factors are removed
exhibiting the feature that the three dimensional structure of a
protein solely depends on its amino-acid ordering information.
This situation gives the hope to some research workers that
protein folding could be reconstructed by computer simulation
\cite{Abe}. But the difficulty of this simulation even for small
proteins is in view.

In this study, the model belonging to the proteins is presented in
section 2. In section 3, the thermodynamical quantities concerning
the variation of the increments of additional enthalpy $\Delta H$
and Gibbs energy $\Delta G$ with respect to temperature have been
expressed and these parameters are related to the parameters of
the proteins with the help of statistical thermodynamics. At the
end, from the curves belonging to the increments of enthalpy and
Gibbs energy, the conclusions relevant to the structures of the
proteins are presented.
\section{Model}
\label{sec:1} A simple model of protein unfolding has been
developed by Bakk \cite{Bakk}. Bakk has started the model by
choosing the water molecules as classical electric dipoles. he has
demanded that non polar dissolution should include important
physical properties in order to model the effect on the folding of
the proteins. In the model used, in the medium of the formed
electric field, the folding and unfolding of the proteins have
been taken as the basis. Here, the electric field is not an
external field but has been used in the modelling of the ice-like
behavior exhibited by the water molecules around the non polar
surfaces. The electric field is a result of the effective behavior
of the non polar dissolvent applied to the protein unfolding.

The modified partition function of the protein system could be
taken as \cite{Bakk}:
\begin{equation}\label{eq:1}
Z(T, E, \mu)=\exp[\exp(\frac{\mu}{k T})z(T, E, N=1)] .
\end{equation}

Here, $\mu$ represents the chemical potential and the term
$\zeta=\exp(\beta\mu)$ corresponding to the pH of the system has
been introduced to the partition function. $z(T,E,N=1)=4 \pi
\exp^{-\frac{\beta b m^2}{2}} \frac{1}{\beta \epsilon_e} {\sinh
\beta \epsilon_e}$ indicates single particle partition function.
Where effective energy is given by $\epsilon_e= \epsilon+ b <m>$
\cite{Bakk}. On the other hand average dipole moment has the form
$<m>=L(\beta \epsilon_e)$.
\section{Variations of the Enthalpy and Gibbs Energy Increments
with Respect to Temperature} Taking electric field $E$ and total
dipole moment $M$ as thermodynamical variables, the first law of
thermodynamics could be written for the proteins in the form:
\begin{equation}\label{eq:2}
    dU=TdS-MdE .
\end{equation}

In this case, change in the enthalpy is;
\begin{equation}\label{eq:3}
    dH=TdS+EdM .
\end{equation}

Enthalpy and entropy increments could also be written by taking
the total differentials of enthalpy $H=H(S,M)$ and entropy
$S=S(T,M)$ which leads to:
\begin{equation}\label{eq:4}
    dH=(\frac{\partial H}{\partial S})_M dS+ (\frac{\partial H}{\partial M})_S dM
\end{equation}
\begin{equation}\label{eq:5}
    dS=(\frac{\partial S}{\partial T})_M dT+ (\frac{\partial S}{\partial M})_T
    dM .
\end{equation}

After substituting Eq.(4) and Eq.(5) in Eq.(3)
\begin{equation}\label{eq:6}
    dH=(\frac{\partial H}{\partial S})_M (\frac{\partial S}{\partial T})_M dT+
[(\frac{\partial H}{\partial S})_M (\frac{\partial S}{\partial
M})_T+ (\frac{\partial H}{\partial M})_S] dM
\end{equation}
has been obtained.

Writing down the expressions $T=(\frac{\partial H}{\partial
S})_M$, $E=(\frac{\partial H}{\partial M})_S$ and $(\frac{\partial
S}{\partial T})_M= \frac{\Delta C_M}{T}$ in Eq.(6) and then
integrating one could get for the enthalpy increment:
\begin{equation}\label{eq:7}
 \Delta H=\int_{T_t}^T C_M dT+\int_{M(T_t)}^{M_(T)}[E+T(\frac{\partial S}{\partial M})_T] dM.
\end{equation}

Using the first law of thermodynamics for the proteins which has
been given by Eq.(2), change in the Gibbs energy
\begin{equation}\label{eq:8}
    dG=-SdT+EdM
\end{equation}
could be written down. With a similar approach, taking the total
differentials of the Gibbs energy $G=G(T,M)$ and total dipole
moment $M=M(T,E)$ and then substituting then in Eq.(8) together
with the expressions $E=(\frac{\partial G}{\partial M})_T$ and
$-S=(\frac{\partial G}{\partial T})_M$ leads after integration to
the expression which gives the Gibbs energy increment:
\begin{equation}\label{eq:9}
 \Delta G=-\int_{T_t}^T S dT+\int_{M(T_t)}^{M_(T)}E dM .
\end{equation}

The free energy of the system has been obtained from the relation
$F=-k T ln Z$ giving:
\begin{equation}\label{eq:10}
 F=-\frac{4 exp(-\beta (\frac{1}{2} b m^2- \mu)) \pi \sinh (\beta \epsilon_e)}{\beta^2
 \epsilon_e} .
\end{equation}

The thermodynamical quantities of the system has been determined
using the entropy relation $S=-(\frac{\partial F}{\partial T})_E$
which in turn leads to:
\begin{eqnarray}\label{eq:11}
 S=\frac{2 exp(-\beta (\frac{1}{2} b m^2- \mu)) k \pi}{\beta \epsilon_e}\times
 \{(-2 \epsilon_e \beta \cosh(\beta \epsilon_e)+ \\
\nonumber
 (4+ \beta b m^2- 2
\beta \mu) \sinh(\beta \epsilon_e)) \}. \\
\nonumber
\end{eqnarray}

The total dipole moment has been calculated from the expression
$M=(\frac{\partial F}{\partial E})_T$ resulting in the following:
\begin{equation}\label{eq:12}
  M=\frac{4 exp(-\beta (\frac{1}{2} b m^2- \mu)) \pi}{\beta^2 \epsilon_e^2}\times
  \{\epsilon_e \beta \cosh(\beta \epsilon_e)- \sinh(\beta \epsilon_e)
  \}.
\nonumber
\end{equation}

The energy of the system is given by $U=-\frac{1}{Z}
\frac{\partial Z}{\partial\beta}$ and when the partition function
is substituted one obtains:
\begin{eqnarray}\label{eq:13}
\nonumber
  U=\frac{exp(-\beta (\frac{1}{2} b m^2- \mu))}{\beta^2
  \epsilon_e}\times \{  (-4 \pi \epsilon_e \beta \cosh(\beta \epsilon_e)+ \\
   2 \pi (2+ \beta b m^2- 2 \beta \mu) \sinh(\beta \epsilon_e))\}. \\
\nonumber
\end{eqnarray}

Additional heat capacity at effective field has been calculated by
making use of the equation $\Delta C_E=-k \beta^2 (\frac{\partial
U}{\partial \beta})_E$ which gives:
\begin{eqnarray}\label{eq:14}
\nonumber
  \Delta C_E=\frac{exp(-\beta (\frac{1}{2} b m^2- \mu)) k \pi}{\beta
  \epsilon_e}\times \{- 8 \epsilon_e \beta \cosh(\beta \epsilon_e)- \\
\nonumber
  (4 \epsilon_e \beta ( \beta b m^2+ 2 \beta \mu)) \cosh(\beta \epsilon_e)+ (8+ \beta (b^2 m^4 \beta- 8 \mu)) \sinh(\beta \epsilon_e)- \\
     \beta 4 b m^2 (\beta \mu -1) \sinh(\beta \epsilon_e)+ 4 \beta^2 (\epsilon_e^2 +\mu^2) \sinh(\beta \epsilon_e) \}. \\
\nonumber
\end{eqnarray}

Using the expression $\Delta C_M- \Delta C_E= - T (\frac{\partial
M}{\partial E})_T [(\frac{\partial E}{\partial T})_M]^2$ which has
been obtained for additional heat capacity at constant total
dipole moment;
\begin{eqnarray}\label{eq:15}
\nonumber
  \Delta C_M=\frac{exp(-\beta (\frac{1}{2} b m^2- \mu)) k \pi}{\beta
  \epsilon_e}\times \{- 4 \epsilon_e \beta (2+ \beta b m^2 -2 \beta \mu) \cosh(\beta \epsilon_e)+ \\
\nonumber
  [8+ \beta (b^2 m^4 \beta -8 \mu -4 b m^2(\beta \mu -1)+ 4 \beta (\epsilon_e^2+ \mu^2))] \sinh(\beta \epsilon_e)- \\
\nonumber
   \frac{k^2 \beta^2}{(-2 \beta \epsilon_e \cosh(\beta \epsilon_e)+
    (2+ \beta^2 \epsilon_e^2) \sinh(\beta \epsilon_e))}\times [\epsilon_e \beta (4 \beta b m^2- 2 \beta \mu) \cosh(\beta \epsilon_e)- \\
     (4+ \beta (b m^2 + 2\beta \epsilon_e- 2 \mu))\sinh(\beta \epsilon_e)]^2 \} \\
\nonumber
\end{eqnarray}
has been calculated.

Another quantity $(\frac{\partial S}{\partial M})_T$ when
calculated gives:
\begin{equation}\label{eq:16}
 (\frac{\partial S}{\partial M})_T=- k \beta \epsilon_e .
\end{equation}
When Eq.(15) and Eq.(16) are substituted in Eq.(7), after
integration gives for the enthalpy increment:

\begin{eqnarray}\label{eq:17}
\nonumber
  \Delta H=\frac{2 \pi exp(-\frac{b m^2 (T_t+ 2 T)}{2 k T_t
  T})}{\epsilon_e^2}\times \{T_t exp(\frac{b m^2 (T_t+ 2 T)+ 2 T \mu}{2 k T_t T})\times  \\
\nonumber
[2 \epsilon_e (b m+ \epsilon- k \epsilon+ (k- 1) \epsilon_e) \cosh(\frac{\epsilon_e}{k T_t})+ k (2 (k- 1) T_t (\epsilon- \epsilon_e)- \\
 \nonumber
 b m (2 T_t+ m \epsilon_e)+ 2 \epsilon_e \mu)\sinh(\frac{\epsilon_e}{k T_t})]+ T exp(\frac{b m^2 T+ T_t \mu}{2 k T_t T})\times  \\
\nonumber
[-2 \epsilon_e (b m+ \epsilon- k \epsilon+ (k- 1) \epsilon_e) \cosh(\frac{\epsilon_e}{k T})+ k (-2 (k- 1) T (\epsilon- \epsilon_e)+ \\
 b m (2 T+ m \epsilon_e)- 2 \epsilon_e \mu) \sinh(\frac{\epsilon_e}{k T})]\ \}.
\end{eqnarray}

Similarly, substituting Eq.(11) into Eq.(9), after integration and
writing down Eq.(12) in this expression, one obtains for the Gibbs
energy increment:
\begin{eqnarray}\label{eq:18}
\nonumber
  \Delta G=\frac{4 k \pi exp(-\frac{b m^2 (T_t+ T)}{2 k T_t
  T})}{\epsilon_e^2}\times \{T_t exp(\frac{\frac{b m^2}{T}+ \frac{2 \mu}{T_t}}{2 k})\times  \\
\nonumber
[-\epsilon \epsilon_e \cosh(\frac{\epsilon_e}{k T_t})+ k T_t (\epsilon +\epsilon_e) \sinh(\frac{\epsilon_e}{k T_t})]-  \\
  \{T exp(\frac{\frac{b m^2}{T_t}+ \frac{2 \mu}{T}}{2 k})\times [-\epsilon \epsilon_e \cosh(\frac{\epsilon_e}{k T})+ k T(\epsilon
+\epsilon_e) \sinh(\frac{\epsilon_e}{k T_t})] \ \}.
\end{eqnarray}

Now the numerical applications of the enthalpy and Gibbs energy
increments will be taken under consideration. The  unfolding and
then folding of the protein molecules take place by a
thermodynamical mechanism. In other words, the original structures
of the macromolecules are determined thermodynamically. In the
following figures, dashed lines represent experimental study where
as solid lines theoretical results. The physical quantities
$\epsilon$, $b$ and $\mu$ of the graphs which have been obtained
in this semi phenomenological theory are determined by fitting to
the experimental results of Privalov
\cite{{Privalov},{Privalov2}}. In Figs. (1) and (2) the variations
of the thermodynamical functions enthalpy and Gibbs energy
increments with respect to temperature are represented for two of
the selected proteins. Yellow (belonging to Myoglobin) and red
(belonging to Ribonuclease) solid lines indicate the variations
with respect to temperature of increments of the thermodynamical
quantities which have been obtained theoretically.

\begin{figure}[t]
\resizebox{0.45\textwidth}{!}{%
  \includegraphics{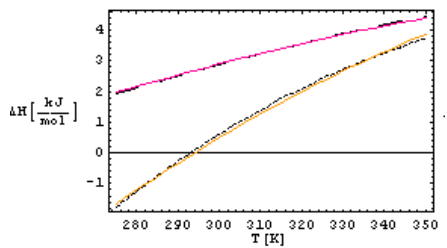}
}
%\vspace{5cm}       % Give the correct figure height in cm
\caption{The variation of the additional enthalpy with
temperature.}
\label{fig:1}       % Give a unique label
\end{figure}

\begin{figure}[t]
\resizebox{0.45\textwidth}{!}{%
  \includegraphics{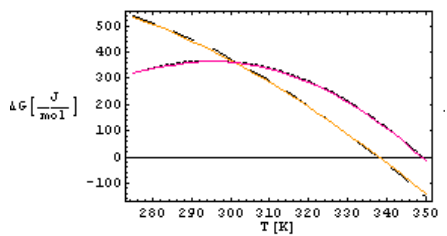}
}
%\vspace{5cm}       % Give the correct figure height in cm
\caption{The variation of the additional Gibbs energy with
temperature.}
\label{fig:2}       % Give a unique label
\end{figure}

\section{Conclusions}
\label{sec} The enthalpy increments $\Delta H$ between the folding
and unfolding states, has a negative value in the cold folding
region. This behavior shows that cold folding has been realized by
giving heat out and this is negative heat capacity, on the other
hand in the $\Delta H>0$ case the system is in the warm folding
region and the heat capacity is positive.

The greatest contribution to the natural structure are provided by
hydrophobic interactions. In an unfolding chain structure,
hydrophobic amino-acid side chains are unfolded by a regular water
molecule lattice and the entropy of the system decreases. But when
the chain undergoes folding and hydrophobic side chains are
collected at intermolecular regions by getting a way from water,
the regular water molecules will be free and as increase in the
total entropy of the system takes place.

In each protein molecule $\Delta G$ Gibbs potential increment, is
a fraction of the quantity $k T_{room}$ in the folding and
unfolding states at each degrees of freedom and $\Delta G$ Gibbs
potential increment is not a monotonic function of temperature.
Gibbs free energy increment between the folding and unfolding
states has a maximum that is $\Delta G> 0$ and thus the unfolded
form is stable and at both sides of this maximum this increment
becomes negative. This means that two transitions take place. One
of them is the unfolding of the proteins at high temperature where
as the other is the folding of the proteins at low temperature.
This phenomenon is known as cold denaturation in other words cold
unfolding. As a result, in the cold denaturation transition hidden
heat is negative and just opposite to this in the warm
denaturation transition hidden heat has a positive value.

During the folding of the protein entropy increases that is to say
in returning to the original structure of the protein it is not
preferred from the point of view of energy. Therefore there must
be some factors which oppose the increase in entropy (i.e.
folding) and encourage on the other hand the folding of the
molecules. Since covalent structure is conserved during unfolding,
forces which oppose the unfolding of the protein could not be
covalent interactions. Transformation of the non polar molecules
into water becomes with a considerable decrease in entropy. The
entropy of hydration of those non polar groups must be temperature
dependent since hydration leads to an important increase in the
heat capacity. The same thing is expected for the enthalpy of
hydration. In fact, if we look at the not hydration effect during
the transition of the non polar components from gas phase to water
we observe that both enthalpy and entropy are negative and the
magnitudes decrease with increasing temperature. Hance, one could
consider that the hydration effect is responsible from the
dissolution of the non polar compounds in water.


%
% BibTeX users please use
% \bibliographystyle{}
% \bibliography{}

\end{document}